\documentclass[aps,prb,twocolumn,superscriptaddress]{revtex4-1}
\usepackage{graphicx,color}
\usepackage{dcolumn}
\usepackage{bm}
\usepackage{amsmath,amsthm,amssymb}
\usepackage{gt15}
\newcommand{\MnFeGe}{Mn$_{1-x}$Fe$_x$Ge}
\newcommand{\FeCoGe}{Fe$_{1-x}$Co$_x$Ge}

\begin{document}

\title{
Doppler shift picture of the Dzyaloshinskii--Moriya interaction
}

\author{Toru Kikuchi}%
\affiliation{%
 RIKEN Center for Emergent Matter Science (CEMS)\\  
2-1 Hirosawa, Wako, Saitama 351-0198, Japan\\}
\author{Takashi Koretsune}
\affiliation{%
 RIKEN Center for Emergent Matter Science (CEMS)\\  
2-1 Hirosawa, Wako, Saitama 351-0198, Japan\\}
\affiliation{%
JST, PRESTO, 4-1-8 Honcho, Kawaguchi, Saitama 332-0012, Japan\\
}
\author{Ryotaro Arita}%
\affiliation{%
 RIKEN Center for Emergent Matter Science (CEMS)\\  
2-1 Hirosawa, Wako, Saitama 351-0198, Japan\\}
\author{Gen Tatara}%
\affiliation{%
 RIKEN Center for Emergent Matter Science (CEMS)\\  
2-1 Hirosawa, Wako, Saitama 351-0198, Japan\\}

\begin{abstract}
We present a physical picture for the emergence of the Dzyaloshinskii--Moriya (DM) interaction
based on the idea of the Doppler shift by an intrinsic spin current induced by spin--orbit interaction under broken inversion symmetry. 
The picture is confirmed by a rigorous effective Hamiltonian theory, which reveals that the DM coefficient is given by the magnitude of the intrinsic spin current. 
The expression is directly applicable to first principles calculations and  
clarifies the relation  between the interaction and the electronic band structures.
Quantitative agreement with experimental results is obtained for the skyrmion compounds
Mn$_{1-x}$Fe$_x$Ge and Fe$_{1-x}$Co$_x$Ge.
\end{abstract}

\date{\today}

\maketitle
\renewcommand{\Jex}{J}

\newpage

Magnets with broken inversion symmetry such as chiral magnets and those in multilayers have been studied intensively in recent years owing to their potential application in nanomagnetic devices.  
Their attractive properties originate from the Dzyaloshinskii--Moriya (DM) interaction\cite{Dzyaloshinskii58,Moriya60}, 
whose Hamiltonian takes the following form in the continuum limit: 
\begin{equation}
H_{\rm DM} \equiv \int d^3r \sum_{ia} D^a_i(\nabla_i\bm n\times \bm n)^a,
\label{DM term}
\end{equation}
with $\nv$ a unit vector denoting the direction of magnetization and $D^a_i$ the DM coefficient.   
While the exchange interaction tends to align the local magnetizations (anti)parallel, the DM interaction makes them twist, which yields numerous magnetization structures at the nanoscale such as helices\cite{Kishine15} and skyrmions\cite{Roszler06, Felser13, Nagaosa13} and gives high mobility to domain walls\cite{Thiaville12, Chen13, Ryu13, Emori13, Torrejon14}.  The DM interaction is also a key ingredient in multiferroics
since it connects magnetizations to electric polarizations\cite{Tokura14, Sergienko06, Katsura05}.
  
In 1960, Moriya\cite{Moriya60} clarified microscopically that the DM interaction arises at the first order of the spin--orbit coupling of electrons.
Since recent investigations involving the DM interaction have become very diverse and precise, its quantitative estimation scheme applicable both to metals and insulators has been strongly demanded.

Katsnelson {\it et al.} calculated the DM interaction by evaluating the energy increase when the magnetization is twisted in a lattice spin model \cite{Katsnelson10}.
Recently, this method has been 
applied to iron borate (FeBO$_3$) to estimate the weak DM interaction ($\sim$0.25 meV) between iron atoms,
and it has been shown that its microscopic expression indeed gives
numerically accurate results\cite{Dmitrienko14t}. 
An evaluation of the twisting energy of the magnetization has been performed also in the continuum spin model \cite{Heide2008,Heide2009}, for example, for Mn$_{1-x}$Fe$_x$Ge and the critical value of $x_c\sim 0.8$ at which the DM interaction changes its sign
has been reproduced successfully\cite{Gayles15}. 
However, 
the relation between the strength or sign of the DM interaction and the electronic band structure 
is not clearly seen in those formalisms based on the twist energy. 

Recently, Berry's phase formalism for the DM interaction
was presented \cite{Freimuth14}, where the relation between the DM interaction and 
the electronic band structure became clear. In this formulation, however,  twist torque operators need to be evaluated, which is not always easy.
It was recently proposed that the DM coefficient is given by a derivative of a spin correlation function with respect to the wave vector \cite{Koretsune15}.
This spin correlation function representation has the advantage of having a direct relation between the DM coefficient and the spin correlation function $\chi_{n\kv}$ for band $n$ and wave vector $\kv$.  However, this approach turned out to be insufficient to reproduce the value $x_c\sim 0.8$ for Mn$_{1-x}$Fe$_x$Ge 
at which the DM coefficient changes its sign \cite{Koretsune15}.

The aim of this paper is to present a new picture for the emergence of the DM interaction as well as to develop a calculation scheme for its coefficient with accuracy and predictability. 
We show that the DM interaction is a consequence of the  \textquotedblleft Doppler shift\textquotedblright\  due to an intrinsic spin current induced by the spin--orbit interaction under broken inversion symmetry.
This fact leads naturally to our main conclusion that the DM coefficient is given by the magnitude of the intrinsic spin current. We also develop a rigorous derivation of the coefficient based on an effective Hamiltonian method.  

Let us start with an intuitive argument.
The spin--orbit interaction with broken inversion symmetry is generally represented by a quantum mechanical Hamiltonian: 
\begin{align}
 H_{\rm so}=& -\sum_{ia}\lambda_i^a \hat{p}_i \sigma_a,
\end{align}
where $\hat{\pv}$ is a momentum operator, $\sigmav$ is a vector of Pauli matrices, and $\lambda_i^a$ is a coefficient specifying the amplitude of the spin--orbit field.
\cite{soform}
In terms of a spin current operator, 
$\hat{j}_{{\rm s},i}^a\equiv \frac{1}{2m}\hat{p}_i\sigma_a$, the interaction is written as 
$H_{\rm so}= -2m \sum_{ia} \lambda_i^a \hat{j}_{{\rm s},i}^a$, 
and it thus generates an intrinsic spin current proportional to $\lambda^a_i$.
The existence of an equilibrium spin current does not contradict the laws of thermodynamics since the current does not do work as far as it is static. 
A similar spin current but with a different origin is known to arise from non-collinear spin structures \cite{TKS_PR08}.

\begin{figure}
        \includegraphics[bb=0 0 1224 738, width=0.4\textwidth]{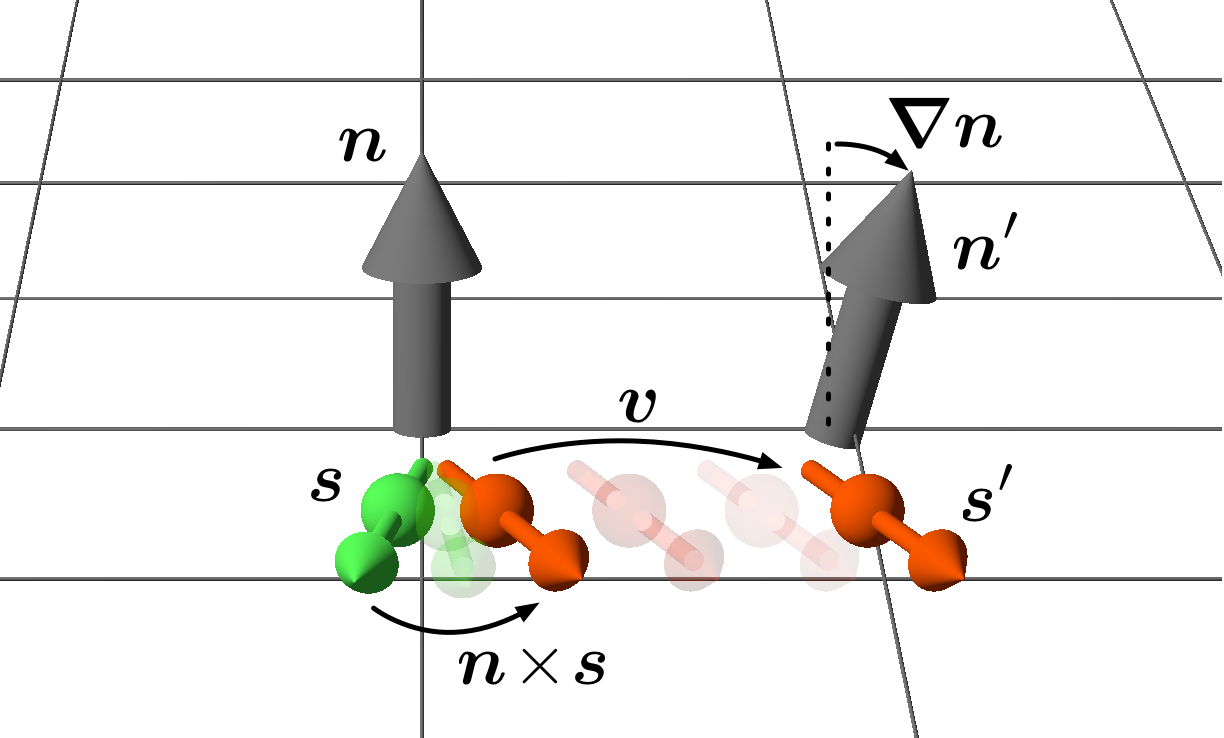}

	\caption{
	Schematic picture showing the mechanism of the spin current Doppler shift. 
	The polarization $\sev$ of the spin current intrinsically determined by the spin--orbit interaction  is deviated because of a torque due to the localized spin $\nv$. This deviation corresponds to the change  of the electron spin's laboratory frame and is described by 
 	a covariant derivative.
 	The effect leads to a  shift in the frequency of the localized spin dynamics, i.e., the Doppler shift. 
	}
	\label{fig:dm_pic}
\end{figure}
We are interested in the magnetic energy of localized spins arising from the electron carrying an intrinsic spin current.
Without the current, the energy is proportional to $(\nabla \nv)^2$ at the second order in the hopping or in the long wavelength limit. 
When a spin current is present, the spatial variation in the localized spins seen from the electron is modified as follows.
The existence of an intrinsic spin current  $j_{{\rm s},i}^a\equiv v_i s_a $  means that the electron with a spin polarization $\sv$ is moving in the direction of the flow $\vv$ (Fig.\ref{fig:dm_pic}). 
This intrinsic spin current is distorted as a result of the {\it sd}-type exchange interaction with the localized spin.
Denoting the localized spin at site $i$ as $\nv$, the torque on the electron spin is proportional to  $\nv\times\sev$, and hence, the electron spin polarization is modified to $\sev'=\sev+\epsilon(\nv\times\sev)$, where $\epsilon$ is a small coefficient.
When the electron with distorted spin hops to a neighboring site having a localized spin direction $\nv'$, it sees the relative direction $\nv'-\sev'$.
Writing $\nv'$ as 
$\nv'\equiv \nv +(\av\cdot\nabla)\nv$ ($\av$ is a vector connecting sites $i$ and $j$ and $\nabla$ represents the discrete derivative), the relative direction is  $\nv'-\sev'=\nv-\sev+(\av\cdot\nabla)\nv+\epsilon(\sev\times\nv)$.
Recalling the fact that $\sev$ is the polarization of the spin current, the above expression indicates that the spatial derivative of the localized spin structure is modified when the electron spin media flow to be a \textquotedblleft covariant derivative\textquotedblright\ as 
\begin{align}
 \mathfrak D_i\nv &= \nabla_i\nv+\eta(\jv_{{\rm s},i}\times\nv), \label{covariant}
\end{align}
where $\eta$ is a coefficient \cite{Kimref}. 
This covariant derivative is interpreted as a result of the Doppler shift as seen from the argument above.  
Only the component of the spin current  perpendicular to the localized spin leads to the Doppler shift, in contrast to the spin-transfer torque effect arising from the adiabatic (parallel)  component. 
Similar Doppler shift for a vector in a moving medium has been known
in the case of the velocity vector of sound wave\cite{LandauLifshitz87}.

When the spatial derivative of the localized spin is modified to the derivative described by Eq.(\ref{covariant}), because of the Doppler shift, the magnetic energy is modified to be proportional to 
$(\mathfrak D_i\nv)^2=(\nabla\nv)^2+2\eta \sum_i\jv_{{\rm s},i}\cdot(\nv\times\nabla_i\nv)+O(\eta^2)$.
We see here that the DM interaction \eqref{DM term} arises and that its magnitude is proportional to that of the intrinsic spin current,
$D_i^a\propto j_{{\rm s},i}^a$.  The Doppler shift interpretation of the DM interaction is natural since the nonreciprocal propagation of spin waves \cite{Kataoka87,Iguchi15} is naturally explained.

The picture presented above is a classical one.  
We present here a rigorous quantum mechanical derivation based on a continuum model, focusing on the metallic case. 
In the field representation, the Hamiltonian is 
\begin{align}
  H&= \intr \sum_\alpha c^\dagger_{\alpha} \biggl[ -\frac{\hbar^2\nabla_i^2}{2m} -\Jex_{\alpha} \nv\cdot\sigmav 
  +\frac{i}{2}\sum_i\lambdav_i\cdot\sigmav \overleftrightarrow{\nabla}_i\biggr] c_{\alpha},
\label{Hamiltonian}
\end{align}
where  $c_\alpha$ and $c^\dagger_\alpha$ are electron creation and annihilation operators for the orbit $\alpha$, respectively, 
$c^\dagger \overleftrightarrow{\nabla}_i c\equiv c^\dagger \nabla_i c- (\nabla_i c^\dagger) c$ and the constant $\lambdav_i$ represents the spin--orbit interaction in the continuum with broken inversion symmetry. 
The local direction of the magnetization $\nv(\rv)$, with $\bm n=(\sin\theta\cos\phi, \sin\theta\sin\phi, \cos\theta)$, is static and $\Jex_{\alpha}$ denotes the exchange constant.  
We consider here a simplified model with the quadratic dispersion and the spin--orbit interaction linear in the momentum but the extension to general cases is straightforward as we shall demonstrate later. 
The effective model (\ref{Hamiltonian}) is derived from a multiband Hubbard model by introducing the magnetization by use of a Hubbard--Stratonovich transformation. 

Considering the case of strong ferromagnets, i.e., large $\Jex_\alpha$, we diagonalize the exchange interaction by introducing a unitary transformation in spin space as 
$c_\alpha(\rv)=U(\rv)a_\alpha(\rv)$, where $U$ is a $2\times2$ unitary matrix satisfying 
$U^\dagger(\nv\cdot\sigmav) U=\sigma_z$ \cite{TKS_PR08}. 
Explicitly, $U$ is chosen as $U=\bm m\cdot \bm \sigma$ with $\bm m\equiv (\sin\frac{\theta}{2}\cos\phi, \sin\frac{\theta}{2}\sin\phi, \cos\frac{\theta}{2})$. 
As a result of the unitary transformation, derivatives of the electron field become covariant derivatives as  $\nabla_i c_\alpha = U(\nabla_i + iA_{{\rm s},i})a_\alpha$ where  
$A_{{\rm s},i} \equiv \sum_a A_{{\rm s},i}^a \frac{\sigma^a}{2} = -iU^\dagger \nabla_i U$ is an SU(2) gauge field, called a spin gauge field, coupling to the electron spin.  
Explicitly, 
$\Av_{{\rm s},i} =\nv\times\nabla_i\nv-A_{{\rm s},i}^z\nv$, where 
$A_{{\rm s},i}^z\equiv (1-\cos\theta)\nabla_i \phi$.
The Hamiltonian for the electron in the rotated frame is $H=H_0+H_A$, where 
\begin{align}
H_0\equiv & \int d^3r  \sum_{\alpha} a^\dagger_{\alpha} \left[- \frac{\hbar^2\nabla_i^2}{2m}  - \Jex_{\alpha} \sigma_z + \frac{i}{2}\sum_i\tilde{\lambdav}_i\cdot \sigmav \overleftrightarrow{\nabla}_i  \right]a_{\alpha},
\label{H0}
\end{align}
with $\tilde{\lambda}^a_i\equiv \sum_b R_{ab}\lambda^b_i$  ($R_{ab}\equiv 2m_a m_b - \delta^{ab}$ being the SO(3) rotation matrix corresponding to $U$)
and 
\begin{align}
H_A \equiv & \int d^3r  \sum_{\alpha}\left[\sum_{ia}\hat{\tilde{j}}_{{\rm s},\alpha,i}^a A^a_{{\rm s},i} +\frac{\hbar^2}{8m}\hat{n}_{{\rm el},\alpha}(A_{{\rm s},i}^a)^2\right].
\label{HA}
\end{align}
Here, 
$\hat{\tilde{j}}_{{\rm s},\alpha,i}^a \equiv - \frac{i\hbar^2}{4m}a^\dagger_\alpha \sigma^a\overleftrightarrow{\nabla}_ia_\alpha - \frac{1}{2}\tilde{\lambda}^a_i \hat{n}_{{\rm el},\alpha}$ is the spin current density operator in the rotated frame and 
$\hat{n}_{{\rm el},\alpha}\equiv a^\dagger_\alpha a_\alpha $.
Equation (\ref{HA}) indicates that the spin gauge field couples to the spin current density.  The information of the magnetization vector $\bm n$ is included in the rotated spin--orbit coupling $\tilde{\lambda}^a_i$ and the spin gauge field $A_{{\rm s},i}^a$.

Our objective is to derive an effective Hamiltonian describing the magnetization by integrating out the conduction electrons.  Here, we are interested in the DM interaction \eqref{DM term}, and it is sufficient to consider the first order derivative terms. 
The effective Hamiltonian is therefore  
\begin{align}
\Heff = & \int d^3r  \sum_{\alpha i a} \tilde{j}_{{\rm s},\alpha,i}^a A^a_{{\rm s},i} ,
\label{Heff}
\end{align}
where $\tilde{j}_{{\rm s},\alpha,i}^a\equiv \average{ \hat{\tilde{j}}_{{\rm s},\alpha,i}^a }$ is the 
expectation value of the spin current density in the rotated frame, which is related to the one in the laboratory as 
${j}_{{\rm s},\alpha,i}^a=\sum_bR_{ab}\tilde{j}_{{\rm s},\alpha,i}^b $. 
By use of the identity 
$\sum_b R_{ab}A_{{\rm s},i}^b = (\nabla_i \bm n\times \bm n)^a + n^aA_{{\rm s},i}^{z}
$,
the effective Hamiltonian reads 
\begin{align}
\Heff = & \int d^3r \lt[\sum_{ia} D_i^a (\nabla_i \bm n\times \bm n)^a +  \sum_i{j}_{{\rm s},i}^\parallel A_{{\rm s},i}^{z} \rt],
\label{Heff2}
\end{align}
where 
${j}_{{\rm s},i}^\parallel \equiv \sum_\alpha \tilde{j}_{{\rm s},\alpha,i}^z=\bm n\cdot\sum_\alpha \bm{j}_{{\rm s},\alpha,i}$, and 
\begin{align}
        D_i^a \equiv
 \sum_{\alpha} {j}_{{\rm s},\alpha,i}^{\perp, a}
\label{DMdef}
\end{align}
with ${j}_{{\rm s},\alpha,i}^{\perp, a}\equiv {j}_{{\rm s},\alpha,i}^{a}-n^a {j}_{{\rm s},\alpha,i}^\parallel $ \cite{spincurrent, spintransfer}.
We see that the DM coefficient is simply given by the 
expectation value
of the spin current density of the conduction electrons.  
Since $\bm n \cdot (\nabla_i\bm n\times \bm n)=0$, only the perpendicular component contributes to the DM interaction, which is  consistent with the intuitive Doppler shift argument (Eq.\eqref{covariant}).

\begin{figure}
        \includegraphics[bb=0 0 1200 600, scale=0.20]{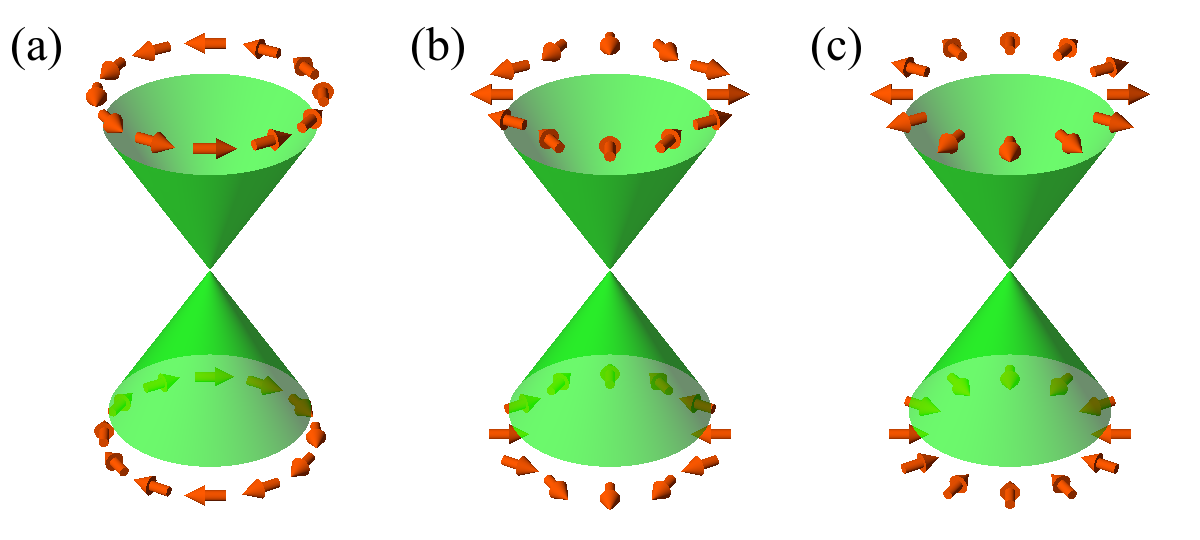}

	\caption{
		Spin texture in momentum space for
		 (a) Rashba, (b) Dresselhaus, and (c) Weyl-type Hamiltonians.
	}
	\label{fig:spin_textures}
\end{figure}

A great advantage of the present formulation for material design is that 
Eq.(\ref{DMdef}) enables the prediction of the DM coefficient based on the features of the band structure.
Let us consider three typical spin configurations of a conduction electron in the momentum space, the Rashba (which arises in polar systems), the Dresselhaus and the Weyl (in chiral systems), represented by the Hamiltonians 
$H_{\rm R}=\alpha(k_x\sigma_y-k_y\sigma_x)$, 
$H_{\rm D}=\beta(k_x\sigma_x-k_y\sigma_y)$ and 
$H_{\rm W}=\gamma(k_x\sigma_x+k_y\sigma_y)$, respectively.
The schematic spin textures are shown in Fig.\ \ref{fig:spin_textures}.
The DM coefficients in those cases (denoted by $D_{\rm R}$, $D_{\rm D}$ and $D_{\rm W}$, respectively) are 
\begin{align}
 &D_{{\rm R},i}^a =  \alpha n_{\rm el} \lt(\begin{array}{ccc} 0 &1&0\\ -1 & 0 & 0 \\ 0&0& 0\end{array}\rt), 
\;
 D_{{\rm D},i}^a = \beta n_{\rm el} \lt(\begin{array}{ccc} 1 &0&0\\ 0 & -1 & 0 \\ 0&0& 0\end{array}\rt),
\;
\notag \\
&D_{{\rm W},i}^a = \gamma n_{\rm el} \lt(\begin{array}{ccc} 1 &0&0\\ 0 & 1 & 0 \\ 0&0& 0\end{array}\rt),
\label{DR and DD and DW}
\end{align}
where $n_{\rm el}$ is the electron density and  the row and column correspond to spatial ($i$) and spin ($a$) indices, respectively.
We therefore see that polar systems
 lead to antisymmetric DM coefficients while diagonal coefficients are expected in non-polar systems, as discussed also in Refs.\ \onlinecite{Kim13, Guengoerdue15} by a different approach.

As for the DM coefficient, our result Eq.(\ref{DMdef}) agrees with that of Ref.\ \onlinecite{Katsnelson10}, derived by evaluating the energy increase when the magnetization is twisted by a local spin rotation.
The expression for the energy increase turned out to be the expectation value of the anticommutator of the spin rotation operator and the hopping matrix element, which is essentially the spin current density. 
On the other hand, the expression discussed in Ref.\ \onlinecite{Koretsune15}, $D\propto \frac{\partial \chi(\qv)}{\partial q}|_{q=0}$ ($\chi$ is the spin correlation function with wave vector $\qv$), and the ones obtained by a Ruderman--Kittel--Kasuya--Yosida (RKKY)  approach\cite{Fert80, Imamura04, Kundu15} are valid when the exchange interaction, $J_{\alpha} \nv\cdot c_\alpha^\dagger\sigmav c_\alpha$ in Eq.\eqref{Hamiltonian}, is weak enough and can be treated perturbatively.  In contrast, in our method that uses a unitary transformation, a strong exchange interaction is assumed.

\begin{figure*}[tbp]
 \begin{minipage}{.49\textwidth}
   \includegraphics[width=85mm]{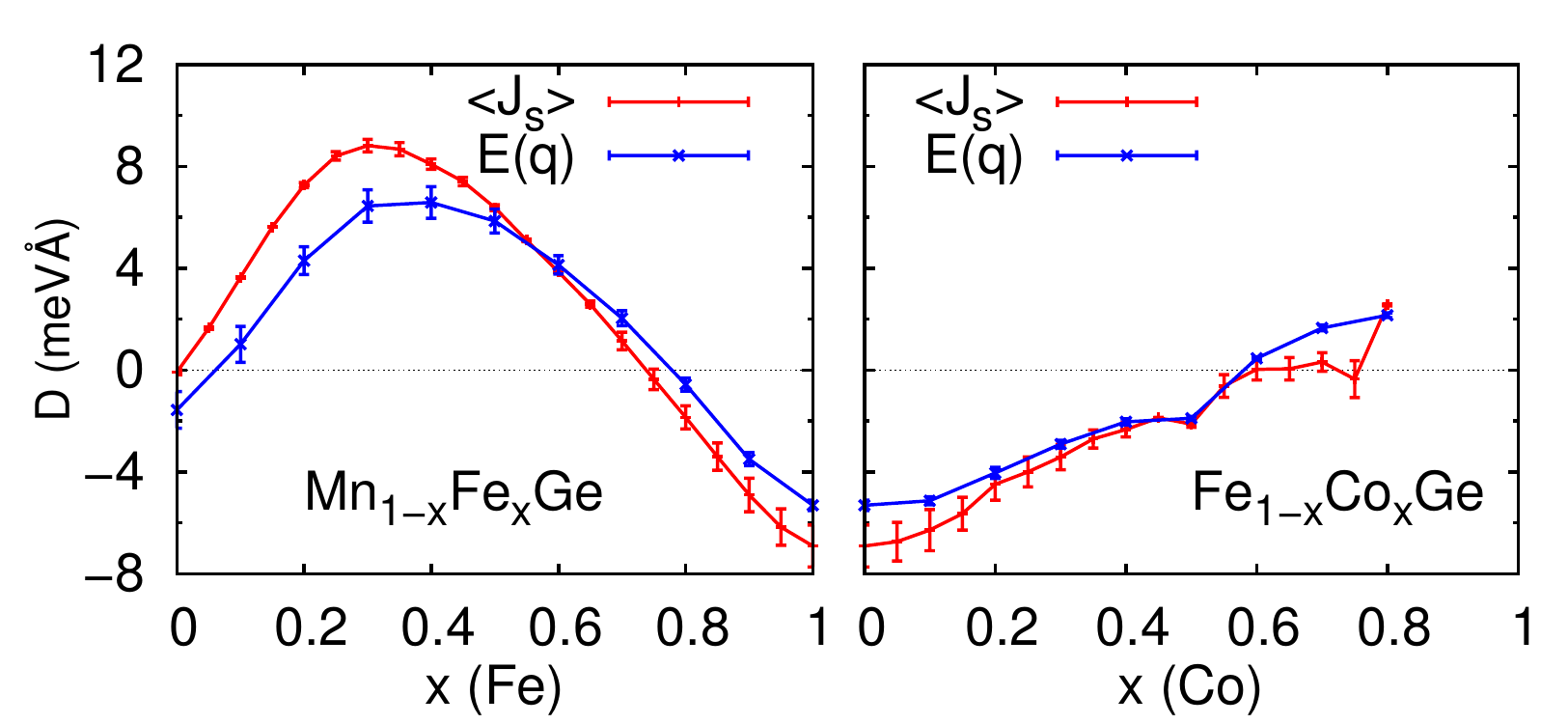}
  \caption{\footnotesize{The DM coefficients $D$ for \MnFeGe\ and \FeCoGe\ calculated using the energies of helical spin structures, $E(q)= Dq+Jq^2$, and as expectation values of the equilibrium spin current $D=\langle \hat j_{\rm s} \rangle$.
		The error bars for each calculation indicate the fitting errors of $E(q)$ and the variances of $D_x^x$ and $D_y^y$, respectively.}}
\label{fig:DM_FeGe}
\end{minipage}
\hfill
 \begin{minipage}{.49\textwidth}
   \includegraphics[width=80mm]{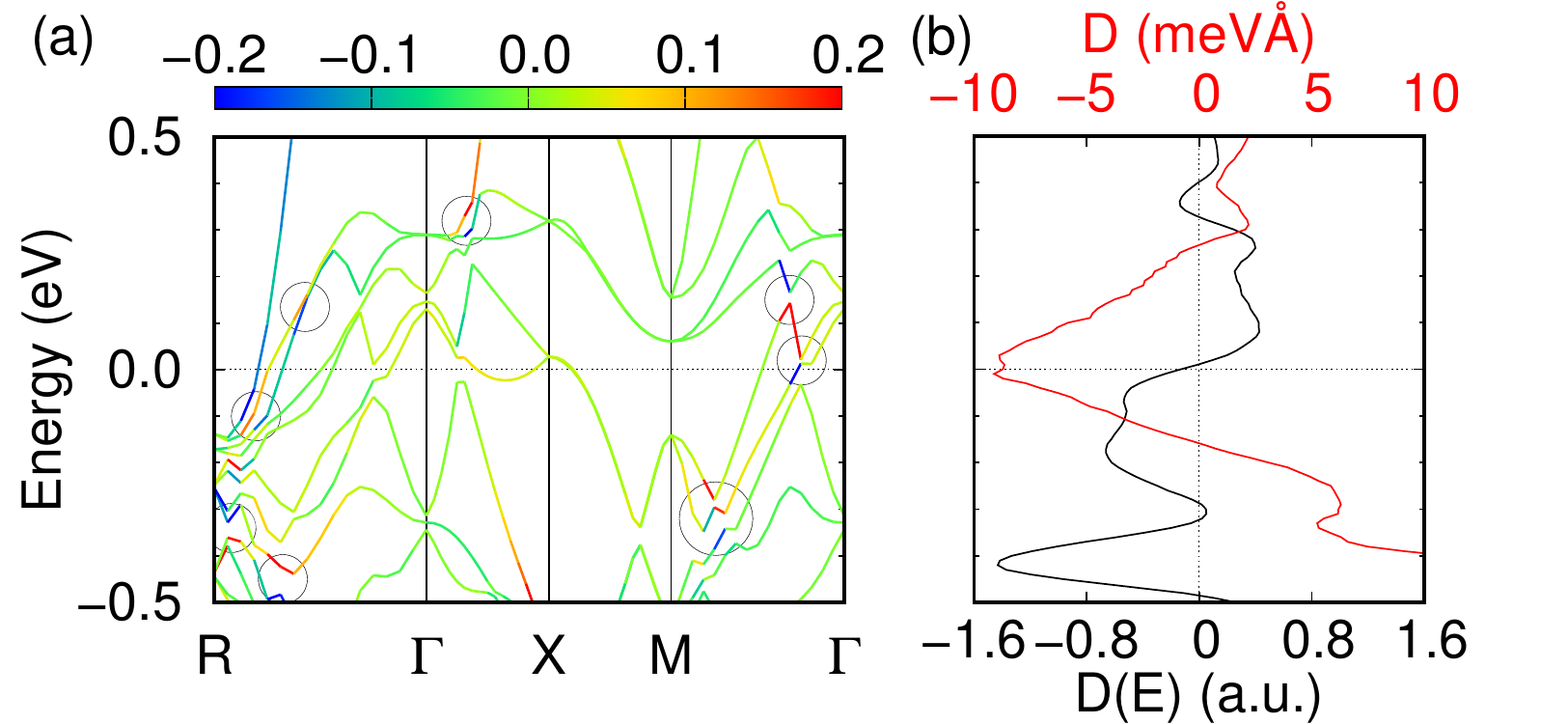}
  \caption{\footnotesize{(a) Contribution of each band to the DM interaction, $D_{n\bm k}$, with dominant band anticrossing points circled, and (b) the energy distribution 
                of the DM interaction, $D(E)$ (black line), 
                for FeGe. 
                The Fermi energy dependence of the DM interaction, $D\equiv\int^{E_F} D(E')f(E')dE'$, within the rigid band approximation is also shown as the red line.}}
  \label{fig:Dk}
 \end{minipage}
\end{figure*}
To examine this approach quantitatively, we perform relativistic electronic-structure calculations for the B20 chiral ferromagnet, FeGe, using quantum-{\sc espresso} code\cite{Giannozzi09}.
We treat exchange and correlation effects within the Perdew--Burke--Ernzerhof (PBE) generalized-gradient approximation\cite{PBE1996} and use ultrasoft pseudopotentials\cite{Vanderbilt1990} with plane-wave cutoffs of 50 Ry for wavefunctions and of 400 Ry for charge densities, respectively.
Brillouin-zone (BZ) integration is carried out on a 8$\times$8$\times$8 $k$-point mesh.
To discuss the atomic composition dependence for Mn$_{1-x}$Fe$_x$Ge and Fe$_{1-x}$Co$_x$Ge, we compute the self-consistent charge densities for several different carrier densities by fixing the atomic geometries and the lattice constant to the experimental values of FeGe.
In all cases, we assume the total magnetic moment in each unit cell to be aligned along the z-axis.

To calculate the DM coefficient using Eq.\ \eqref{DMdef} from the first principles, we consider a general form of the spin current density as follows:
\begin{align}
	j_{{\rm s},\alpha,i}^a = \sum_{\bm k}\frac{1}{4} \langle  c_{\bm k \alpha}^\dagger ( v_{i} \sigma_a + \sigma_a v_{i} ) c_{\bm k \alpha} \rangle,
	\label{J_DFT}
\end{align}
where the velocity operator is defined as $v_i = d H_{\bm k}/ d k_i$ with 
$H_{\bm k}=e^{-i\bm k\cdot \bm x}H e^{i\bm k\cdot \bm x}$.
Due to the symmetry of B20 magnets, we focus on $D_x^x=\sum_\alpha j_{{\rm s},\alpha,x}^x$ and $D_y^y$.
In fact, the other components of $D_i^a$ are found to be negligible compared to $D_x^x$ and $D_y^y$.
For the BZ integration in Eq.\ \eqref{J_DFT}, we employ the Wannier interpolation technique\cite{Marzari1997,Souza2001,Mostofi2008} with Fe 3d and Ge 4p orbitals.
The calculations below are performed on a 32$\times$32$\times$32 $k$-point mesh.
We confirmed that the results on a 64$\times$64$\times$64 $k$-point mesh do not differ much from the results of 32$\times$32$\times$32 $k$-point calculations.

For comparison, we also calculate the DM coefficient using the energies of helical spin structures\cite{Heide2008,Heide2009}.
We assume the helical spin moment to be $\bm M_{\bm q}(\bm r) = M (\cos(\bm q \cdot \bm r), \sin(\bm q \cdot \bm r), 0)$ with $\bm q = (0, 0, q)$ using the generalized Bloch theorem\cite{Sandratskii1986} and calculate the energies of electronic structures, $E(q)$, by the VASP code\cite{VASP,VASP_noncol} within the PBE generalized-gradient approximation.
We use projector augmented wave (PAW) pseudopotentials with a plane-wave cutoff of 500 eV and a 10$\times$10$\times$10 $k$-point mesh.  
Within the continuum model, the energies of the helical spin structures can be easily obtained as $E(q) = D q + J q^2$.
Thus, by extracting the first order in $q$, we can estimate the DM coefficient.

Figure \ref{fig:DM_FeGe} shows the DM coefficients for Mn$_{1-x}$Fe$_x$Ge and Fe$_{1-x}$Co$_x$Ge obtained using the two approaches.
The result obtained by evaluating $E(q)$ for Mn$_{1-x}$Fe$_x$Ge is consistent with Ref.\ \onlinecite{Gayles15} around $x=1$, while the values are slightly different around $x=0$.
This is because Ref.\ \onlinecite{Gayles15} uses the virtual crystal approximation and $x$-dependent crystal structures.
As can be seen, the results of the two methods agree well with each other.
Furthermore, both calculations well reproduce the position of the sign change observed in experiments, $x_c=0.8$ for Mn$_{1-x}$Fe$_x$Ge\cite{Shibata2013,Grigoriev13} and $x_c=0.6$ for Fe$_{1-x}$Co$_x$Ge\cite{Grigoriev2014}.


In our approach, it is easy to discuss the relationship between the DM interaction and the band structure.
In fact, we can rewrite Eq.\ \eqref{DMdef} and Eq.\ \eqref{J_DFT} as
\begin{align}
	D = \sum_{n \bm k} D_{n\bm k} f(\epsilon_{n\bm k}) = \int D(E) f(E) dE,
\end{align}
where $n$ is the band index and $f(E)$ is the Fermi distribution function.
The first equation defines the contribution of each band to the DM interaction, $D_{n\bm k}$, and the second equation defines the density of the DM interaction, $D(E)$.
Figure \ref{fig:Dk} (a) shows $D_{n\bm k}$ for the band structure of FeGe.
As discussed before\cite{Koretsune15}, we can find that the band anticrossing points are important for the DM interaction.
The density of the DM interaction, $D(E)$, shown in Fig.\ \ref{fig:Dk} (b), also gives useful information to discuss the carrier density dependence of the DM interaction. 
That is, in this case, $D(E) < 0$ for $E<0$ and $D(E) > 0$ for $E>0$ indicate the dip structure around FeGe ($E=0$) and the resulting two sign changes in \MnFeGe\ and \FeCoGe.

In summary, we have shown that the origin of the Dzyaloshinskii--Moriya (DM) interaction is the Doppler shift due to an intrinsic spin current induced by the spin--orbit interaction under broken inversion symmetry.
The idea was confirmed by a rigorous effective Hamiltonian approach, and an {\it ab initio} formalism to calculate the DM constant with quantitative accuracy was developed.  
Our identification of the DM constant as the spin current density  will be useful for analyzing multilayered systems\cite{Yang15, Schweinghaus16, Yang16} with high spatial resolution.
Extensions of our formalism to the non-equilibrium cases such as under strain and voltage \cite{Shibata15, Chacon15, Nawaoka15} 
are important future directions.

\acknowledgements
The authors thank H. Fukuyama, H. Kawaguchi, H. Kohno, N. Nagaosa, M. Ogata, S. Seki and Y. Suzuki for valuable discussions.
This work was supported by a Grant-in-Aid for Scientific Research (C) (Grant No. 25400344) from the Japan Society for the Promotion of Science and  
a Grant-in-Aid for Scientific Research on Innovative Areas (Grant No.26103006) from The Ministry of Education, Culture, Sports, Science and Technology (MEXT), Japan.  The work of T. Koretsune was supported by JST, PRESTO.



\bibliographystyle{apsrev4-1}
\bibliography{Kikuchiref}

\begin{thebibliography}{52}
\bibitem [{sof()}]{soform}%
  \BibitemOpen
  \href@noop {} {}\bibinfo {note} {The spin--orbit interaction derived from the
  Dirac equation corresponds to $\lambda_i^a=\sum_j \epsilon_{ija}E_j$, where
  $\epsilon_{ija}$ is the totally antisymmetric tensor and $E_j$ denotes an
  electric field.}\BibitemShut {Stop}%
\bibitem [{Kim()}]{Kimref}%
  \BibitemOpen
  \href@noop {} {}\bibinfo {note} {Similar expression for the covariant
  derivative has been discussed for the particular case of the Rashba
  spin--orbit interaction in Ref.\ \onlinecite{Kim13}, but it was not
  associated with the intrinsic spin current.}\BibitemShut {Stop}%
\bibitem [{spi({\natexlab{a}})}]{spincurrent}%
  \BibitemOpen
 \bibinfo {note} {The spin current in the
  expression of Eq.(\ref{DMdef}) is the intrinsic part induced by the
  spin--orbit interaction in the absence of magnetization structures.
  Obviously, the spin current generated by the magnetization structure, ${\bm
  j}_{{\rm s},\alpha,i}\propto (\partial_i \bm n\times \bm n)$, results in only
  the terms with higher order derivatives in the effective
  Hamiltonian.}\BibitemShut {Stop}%
\bibitem [{spi({\natexlab{b}})}]{spintransfer}%
  \BibitemOpen
  \bibinfo {note} {As in seen in Eq.
  (\ref{Heff2}), spin transfer effect arises if the adiabatic contribution
  ${j}_{{\rm s},i}^\parallel$ remains finite, but such spontaneous motion does
  not arise at least when the spin--orbit interaction is weak. We have checked
  that the adiabatic spin current vanishes to the third order of $\lambda$ in
  our model.}\BibitemShut {Stop}%
\end{thebibliography}

\end{document}